\documentclass[twocolumn,showpacs,prl]{revtex4}

\usepackage{graphicx}% Include figure files
\begin{document}

\title{Qutrit state engineering with biphotons}
\author{Yu.I.Bogdanov}

\affiliation{Russian Control System Agency,
"Angstrem", Moscow 124460 Russia.}
\author{M.V.Chekhova, S.P.Kulik, G.A.Maslennikov,
A.A.Zhukov.}
\affiliation{Department of Physics, Moscow M.V.
Lomonosov State University, 119992 Moscow, Russia.}
\email{postmast@qopt.phys.msu.su}
\author{C.H.Oh, M.K.Tey}
\affiliation{Department of Physics, Faculty of Science, National
University of Singapore, 117542 Singapore.}
\date{\today}
\begin{abstract}
The novel experimental realization of three-level optical quantum
systems is presented. We use the polarization state of biphotons
to generate a specific sequence of states that are used in the
extended version of BB84 QKD protocol. We experimentally verify
the orthogonality of the basic states and demonstrate the ability
to easily switch between them. The tomography procedure is
employed to reconstruct the density matrices of generated states.
\end{abstract}
\maketitle
 The art of quantum state engineering, i.e., the ability
to generate, transmit and measure quantum systems is of great
importance in the emerging field of quantum information
technology. A vast majority of protocols relying on the properties
of two-level quantum systems (qubits) were introduced and
experimentally realized. But naturally, there arose a question of
an extension of dimensionality of systems used as information
carriers and the new features that this extension can offer. The
simplest extension provokes the usage of three-state quantum
systems (qutrits). Recently new quantum key distribution (QKD)
protocols were proposed that dealt specifically with qutrits
\cite{peres:00,kasz:03} and the eavesdropping analysis showed that
this systems were more robust against specific classes of
eavesdropping attacks \cite{bruss:02,durt:03}. The other advantage
of using multilevel systems is their possible implementation in
the fundamental tests of quantum mechanics \cite{collins:02},
giving more divergence from classical theory. The usage of
multilevel systems also provides a possibility to introduce very
specific protocols, which cannot be implemented with the help of
qubits such as Quantum Bit Commitment, for example \cite{lang:04}.
Recent experiments on realization of qutrits rely on several
issues. In one case, the interferometric procedure is used, where
entangled qutrits are generated by sending an entangled photon
pair through a multi-armed interferometer \cite{rob:01}. The
number of arms defines the dimensionality of the system. Other
techniques rely on the properties of orbital angular momentum of
single photons \cite{lang:04,zei:01,zei:03} and on postselection
of qutrits from four-photon states \cite{antia:01}. Unfortunately
all mentioned techniques provide only a partial control over a
qutrit state. For example in a method, mentioned in
\cite{lang:04,zei:01,zei:03} a specific hologram should be made
for given qutrit state. The real parts of the amplitudes of a
qutrit, generated in \cite{rob:01} are fixed by a characteristics
of a fiber tritter, making it hard to switch between the states.
Besides, in this method no tomographic control over generated
state had been yet performed.
\par
\begin{figure}[!ht]
\includegraphics[width=0.35\textwidth]{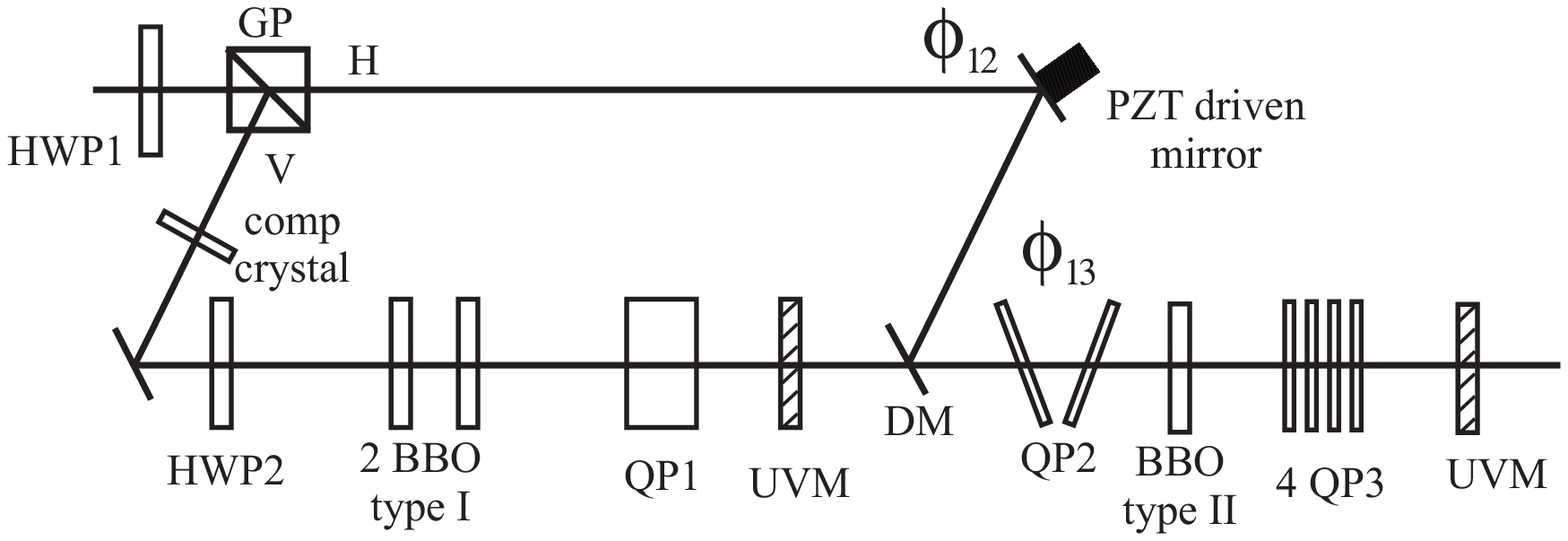}
\caption{Experimental setup (preparation part)}\label{setup1}
\end{figure}

\begin{figure}[!ht]
\includegraphics[width=0.25\textwidth]{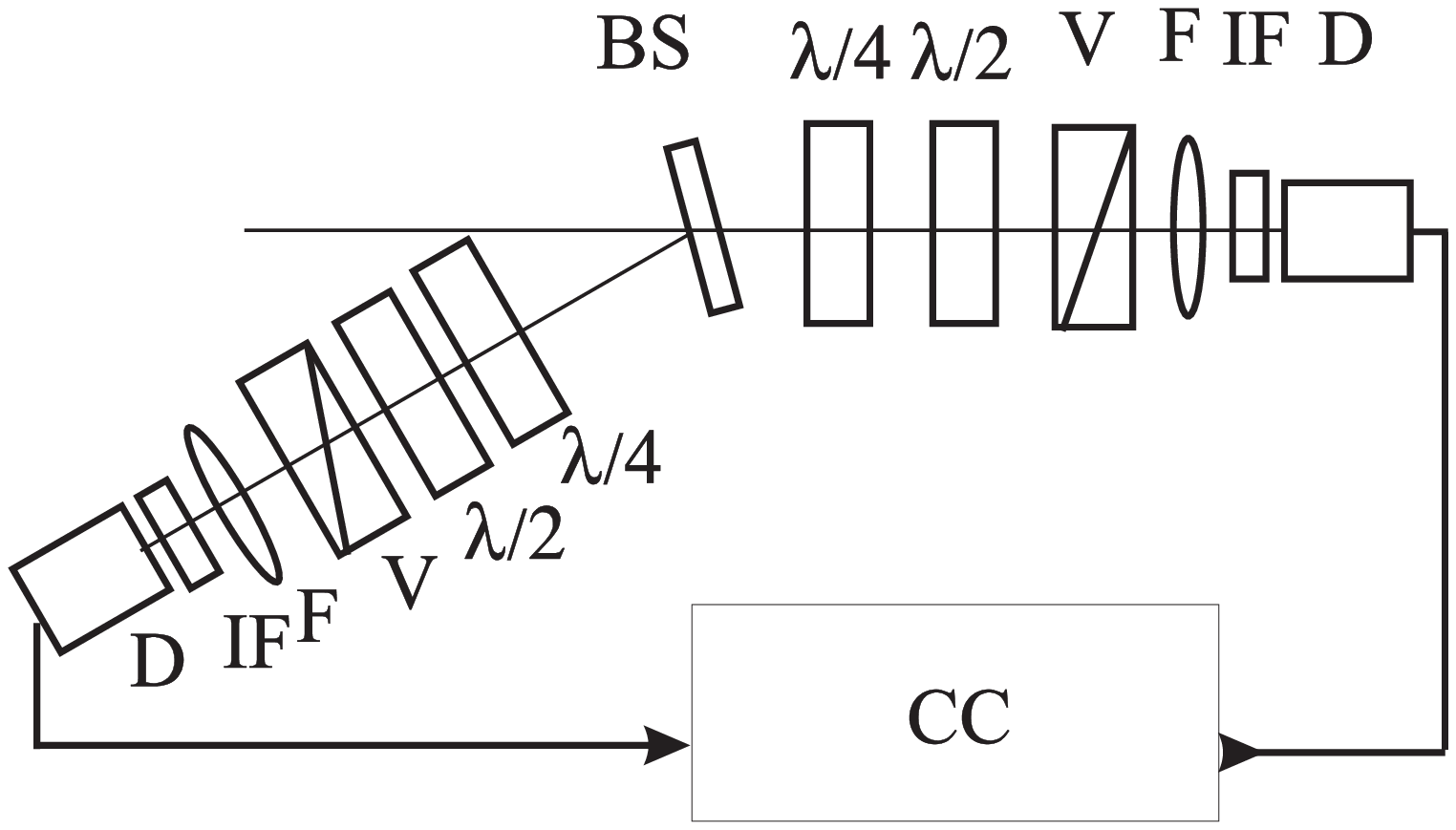}
\caption{Experimental setup (measurement part)} \label{setup2}
\end{figure}

 In this paper we
report the experimental realization of arbitrary qutrit states
that exploits the polarization state of single-mode biphoton
field. This field consists of pairs of correlated photons, is most
easily obtained with the help of spontaneous parametric
down-conversion (SPDC). By saying "single-mode" we mean that twin
photons forming a biphoton have equal frequencies and propagate
along the same direction. A pure polarization state of such field
can be written as the following superposition of three basic
states.
\begin{equation}\footnotesize
|\Psi\rangle=c_{1}|2,0\rangle+c_{2}|1,1\rangle+c_{3}|0,2\rangle = c_{1}|\alpha\rangle+c_{2}|\beta\rangle+c_{3}|\gamma\rangle,       \label{eq:state}
\end{equation}
where $c_i=|c_i|e^{i\phi_i}$ are complex probability amplitudes.
The states $|2,0\rangle$ and $|0,2\rangle$ correspond to type I
phase-matching where twin photons have collinear polarization
vectors (for example, state $|2,0\rangle$ corresponds to two
photons being in horizontal $H$ polarization mode), and state
$|1,1\rangle$ is obtained via type II phase-matching, where
photons are polarized orthogonally (say, one of them is in $H$ and
the other one is in $V$ mode). There exists an alternative
representation of state $|\Psi\rangle$ that maps the state onto
the surface of the Poincare sphere \cite{masha:01} \vspace{0.5cm}
\begin{equation}\footnotesize
|\Psi\rangle=\frac{a^\dagger(\theta,\phi)a^\dagger(\theta^\prime,\phi^\prime)|vac\rangle}{\parallel
a^\dagger(\theta,\phi)a^\dagger(\theta^\prime,\phi^\prime)|vac\rangle
\parallel}, \label{eq:state2}
\end{equation}
where $a^\dagger(\theta,\phi)$ and
$a^\dagger(\theta^\prime,\phi^\prime)$ are the creation operators
of a photon in a certain polarization mode $a^\dagger(\theta,\phi)
= cos(\theta/2)a^\dagger+e^{(i\phi)}sin(\theta/2)b^\dagger$,
$a^\dagger,b^\dagger$ are photon creation operators in
correspondingly horizontal and vertical polarization modes,
$\theta\in[0,\pi]$, $\phi\in[0,2\pi]$ are polar and azimuthal
angles that define the position of each photon on the surface of a
sphere. The values of the angles can be calculated using the
amplitudes and the phases of $c_{i}$. The operational
orthogonality criterion for the polarization states of single-mode
biphotons was proposed in \cite{masha:02} and experimentally
verified in \cite{leo:04}. According to the orthogonality
criterion for biphoton polarization states, two polarization
states $\Psi_{a}$ and $\Psi_{b}$ are orthogonal if one observes
zero coincidence rate in the Brown-Twiss scheme, provided that the
state $\Psi_{a}$ is at the input, and polarization filters in each
arm are tuned to assure maximal transmittance of each photon
forming the state $\Psi_{b}$ (set state). The goal of our work was
to demonstrate the ability to prepare any given polarization state
$|\Psi\rangle$ and as a straightforward and practical example of
given states, we chose the specific sequence that was presented in
\cite{peres:00}. This sequence of 12 states forms four mutually
unbiased bases with three states in each, and can be used in an
extended version of BB84 QKD protocol for qutrits. The 12 states
are defined in Table I.
\begin{center}
\begin{table}[!ht]
\begin{tabular}{|c|c|c|c|c|c|c|}\hline
State&$|c_1|$&$|c_2|$&$|c_3|$&$\phi_1$&$\phi_2$&$\phi_3$\\\hline
$|\alpha\rangle$&$1$&$0$&$0$&$0$&$0$&$0$\\
\hline
$|\beta\rangle$&$0$&$1$&$0$&$0$&$0$&$0$\\
\hline
$|\gamma\rangle$&$0$&$0$&$1$&$0$&$0$&$0$\\
\hline
$|\alpha^\prime\rangle$&$\frac{1}{\sqrt{3}}$&$\frac{1}{\sqrt{3}}$&$\frac{1}{\sqrt{3}}$&$0$&$0$&$0$\\
\hline
$|\beta^\prime\rangle$&$\frac{1}{\sqrt{3}}$&$\frac{1}{\sqrt{3}}$&$\frac{1}{\sqrt{3}}$&$0$&$120^\circ$&$-120^\circ$\\
\hline
$|\gamma^\prime\rangle$&$\frac{1}{\sqrt{3}}$&$\frac{1}{\sqrt{3}}$&$\frac{1}{\sqrt{3}}$&$0$&$-120^\circ$&$120^\circ$\\
\hline
$|\alpha^{\prime\prime}\rangle$&$\frac{1}{\sqrt{3}}$&$\frac{1}{\sqrt{3}}$&$\frac{1}{\sqrt{3}}$&$120^\circ$&$0$&$0$\\
\hline
$|\beta^{\prime\prime}\rangle$&$\frac{1}{\sqrt{3}}$&$\frac{1}{\sqrt{3}}$&$\frac{1}{\sqrt{3}}$&$0$&$120^\circ$&$0$\\
\hline
$|\gamma^{\prime\prime}\rangle$&$\frac{1}{\sqrt{3}}$&$\frac{1}{\sqrt{3}}$&$\frac{1}{\sqrt{3}}$&$0$&$0$&$120^\circ$\\
\hline
$|\alpha^{\prime\prime\prime}\rangle$&$\frac{1}{\sqrt{3}}$&$\frac{1}{\sqrt{3}}$&$\frac{1}{\sqrt{3}}$&$-120^\circ$&$0$&$0$\\
\hline
$|\beta^{\prime\prime\prime}\rangle$&$\frac{1}{\sqrt{3}}$&$\frac{1}{\sqrt{3}}$&$\frac{1}{\sqrt{3}}$&$0$&$-120^\circ$&$0$\\
\hline
$|\gamma^{\prime\prime\prime}\rangle$&$\frac{1}{\sqrt{3}}$&$\frac{1}{\sqrt{3}}$&$\frac{1}{\sqrt{3}}$&$0$&$0$&$-120^\circ$\\
\hline
\end{tabular}
\caption{12 states used in qutrit QKD protocol}
\end{table}
\end{center}
    The preparation part of our setup (Fig.~\ref{setup1}) is built on the base of a balanced Mach-Zehnder interferometer
    (MZI) \cite{gleb:03}.
    The pump part consists of frequency doubled "Coherent Mira 900" femtosecond laser,
    operated at central wavelength of 800 nm, 75 MHz repetition rate and with a pulse width of 100 fs, average pump power
    was 20 mW. The Glan-Tompson
    prism (GP), transmitting the horizontally polarized fraction of the UV pump and reflecting the vertically polarized
    fraction, serves as an input mirror of MZI. The reflected part, after passing the compensation BBO crystal and a
    half-wave plate (HWP2), pumps two consecutive 1 mm thick type-I BBO crystals whose optical axis are oriented
    perpendicularly with respect to each other. The biphotons from these crystals pass through a 10 mm quartz
    plate (QP1) that serves as a compensator, and the pump is reflected by an UV mirror. Then the biphotons arrive at a dichroic
    mirror (DM) that is designed to transmit them and to reflect the horizontally polarized component of the pump
    coming from the upper arm of MZI. A piezoelectric translator (PZT) was used to change the phase shift of
    the horizontal component of the pump with respect to the one propagating in the lower arm.
    The UV beam, reflected from DM serves as a pump for 1 mm thick type-II BBO crystal.
     Two 1 mm quartz plates (QP2) can be
    rotated along the optical axis to
    introduce a phase shift between horizontally and vertically polarized type-I biphotons, and a set of four 1 mm thick
    quartz plates (QP3) serves to compensate the group velocity delay between orthogonally polarized
    photons during their propagation in type II BBO crystal. The measurement setup (Fig.~\ref{setup2}) consists
    of a Brown-Twiss scheme with a non-polarizing 50/50 beamsplitter; each arm contains consecutively placed quarter- and
    half waveplates and an analyzer that was set to transmit the vertical polarization.
    This sequence of waveplates and
    analyzer is referred to as a polarization filter. Interference filters of 5 nm bandwidth, centered at 800 nm and pinholes
    are used for spectral and spatial modal selection of biphotons. We use EGG-SPCM-AQR-15 single photon counting
    modules as our detectors (D1 and D2). We should mention, that due
    to the low pump power, the stimulated processes in our setup are negligibly small and only pairs of photons
    have been generated.
    The measurement of the generated states is done using the tomography protocol that was developed for polarization
    qutrits \cite{bogdan:03a}. In order to reconstruct the density matrix of the measured state (which is generally mixed)
    one has to perform nine projective measurements of the fourth-order moments of the field for different settings
    of polarization filters. Polarization density matrix can be defined in the following way in
    terms of these moments \cite{dnk:97, bogdan:03a}.
\begin{center}\footnotesize
\begin{equation}
\begin{array}{cc}
2\rho_{11}=\langle a^{\dagger^2}a^2\rangle, &
\sqrt{2}\rho_{21}=\langle a^{\dagger^2}ab\rangle,\\
2\rho_{33}=\langle b^{\dagger^2}b^2\rangle, &
\sqrt{2}\rho_{32}=\langle a^\dagger b^\dagger b^2\rangle,\\
\rho_{22}=\langle a^\dagger b^\dagger a b\rangle, &
2\rho_{31}=\langle a^{\dagger^2} b^2\rangle.
\end{array}
\label{eq:rho}
\end{equation}
\end{center}
This configuration of the measurement setup (Fig.~\ref {setup2})
allows us to verify the orthogonality of the states that belong to
the same basis.

   \textbf{Compensation}. In order to have the three terms in superposition (\ref{eq:state}) interfering, one must achieve their
    perfect overlap in frequency, momentum and time domains. From the experimental point of view this means that
    the biphoton wavepackets coming from the two type I crystals and from the type II crystal must be overlapped.
    The overlap in the frequency domain is achieved by the usage of 5nm bandwidth interference filters and the overlap
    in momentum is ensured by using pinholes that select one spatial mode of the biphoton field. But the overlap
    in time cannot be achieved easily when using a pulsed laser source, because it is necessary to compensate for all
    the group delays that biphoton wavepackets acquire during their propagation through the optical elements of the
    setup \cite{sergei:00}. It was found that in order to overlap type-I biphotons with type-II, the pump pulse from
    the lower arm must be delayed. In our case the value of the delay is ~50 ps.
    This was achieved by inserting an additional 2 mm BBO crystal in the lower arm.
    The overlap between the states $|2,0\rangle$ and $|0,2\rangle$ was achieved by inserting a 10 mm quartz plate
    directly after the two type I BBO crystals. After overlapping the biphotons with these techniques, the average
    coincidence count rate that we observed was of about 1 Hz. The high visibility of interference patterns that we obtained
    was a criterion for a good compensation.

   \textbf{Experimental procedure}. In order to create a given qutrit state we needed to have independent control over four real parameters -
    two relative amplitudes and two relative phases. In the experiment we used HWP1 to control the amplitude of the
    state $|1,1\rangle$, and HWP2 to control the relative amplitudes of the states $|2,0\rangle$ and $|0,2\rangle$.
    The relative phase $\phi_{13}=\phi_{3}-\phi_{1}$ between the states $|2,0\rangle$ and $|0,2\rangle$ can be controlled
    with the help of rotating quartz plates (QP2). The relation of the phase $\phi_{12}=\phi_{2}-\phi_{1}$ between
    the state $|\Psi^\prime\rangle=|2,0\rangle+e^{i\phi_{13}}|0,2\rangle$ and $|1,1\rangle$ to the voltage applied to PZT can be found by monitoring the pump
    interference pattern in M-Z interferometer. We found that the change of voltage by 1 V resulted in the phase shift of
    $51.7^\circ$ and $\phi_{12}$ grew linearly with the applied voltage.
    \par States that constitute the first basis are trivial (Table I). They can be produced with the help of a single
    crystal, corresponding to type I or type II interaction. State $|2,0\rangle$ is generated when first $\lambda/2$ (HWP1) angle corresponds to the maximal reflection
    of the pump beam into the lower arm of a Mach-Zehnder and the angle of the second half-lambda waveplate (HWP2) is equal to $0^{\circ}$.
    In order to generate state $|0,2\rangle$, the HWP2 must be rotated by $45^{\circ}$ degrees from $0^{\circ}$,
    and to generate state $|1,1\rangle$ the HWP1 is rotated such, that the whole pump goes into the upper arm of Mach-Zehnder.
    Therefore, in the following, we will consider only the generation of the rest nine states, i.e. those forming the other three bases.
    According to Table I, only the relative phases
    between the basic states are to be varied. This allows us to use the same settings of the HWP's for the
    generation of nine states. It is also convenient to perform three sets of data acquisition - for the fixed
    $\phi_{13}$ values of $0$, $+120^\circ$ and $-120^\circ$, we change $\phi_{12}$ values in the range of, say, few
    periods and perform all tomographic measurements for each value of the phase $\phi_{12}$. Then we select the values
    of $\phi_{12}$ that correspond to the generation of the required state. For example, in order to generate the state
    $\beta^\prime$, we use $\phi_{13}=-120^\circ$ and $\phi_{12}=120^\circ$. The values of the moments at this point
    allow us to restore a raw density matrix of the generated state and compare it to the theoretical value.\par
    The following procedure was used in order to verify the orthogonality of the states that form a certain basis.
    First we chose a set state to which we would tune our polarization filters. Then the values of the angles of
    quarter- and half- waveplates (Fig. ~\ref{setup2}) ($\chi_1,\theta_1,\chi_2,\theta_2$) that assure the maximal projection of the polarization state
    of each photon on the $V$ direction can be calculated by mapping the set state on the Poincare
    sphere. Here, the lower index "1" corresponds to the transmitted
    arm, and the index "2" to the reflected arm of BS
    We chose states $|\alpha^\prime\rangle$, $|\alpha^{\prime\prime}\rangle$ and $|\alpha^{\prime\prime\prime}\rangle$
    to be our set states for each basis. Then, by setting the phase $\phi_{13}$ fixed and by varying the
    phase $\phi_{12}$ we measured the number of coincidence counts that correspond to the certain fourth order moment of
    the field. According to the orthogonality criterion, the coincidence rate should fall to zero when the values
    of $\phi_{13}$ and $\phi_{12}$ correspond to the generation of the states orthogonal to the set ones.

\begin{figure}[!ht]
\includegraphics[width=0.32\textwidth]{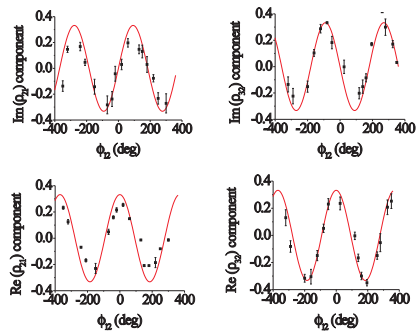}
\caption{Imaginary and real values of non-diagonal density matrix
components used to reconstruct state
$|\beta^{\prime\prime}\rangle$. Theoretical dependence is plotted
with a solid curve.} \label{Res}
\end{figure}

  \textbf{Results  and discussion}. Let us consider the generation of the state $|\beta^{\prime\prime}\rangle$. In this case
    $\phi_{13}=0, \phi_{12}=120$. In Fig. ~\ref{Res} the measured values of the real and imaginary parts of the density matrix
    components $\rho_{21}$ and $\rho_{32}$ on phase $\phi_{12}$ are shown as function of the phase
    $\phi_{12}$. The number of accidental coincidences was negligibly
    small and was not subtracted in data processing.

    The phase $\phi_{13}=0$ remained constant during the tomography procedure. After obtaining
    the dependence of the moments $\rho_{21}$ and $\rho_{32}$ on phase $\phi_{12}$ we fitted our data with
    theoretical dependencies, using the least-square approximation method. The obtained values of
    all components were substituted in Eq.~\ref{eq:rho}. The obtained density matrix for state $|\beta^{\prime\prime}\rangle$
    is given below.

\begin{center}\footnotesize
\begin{equation}
\rho_{\beta^{\prime\prime}} =
\left( \begin{array}{ccc}
0.355&-0.054-0.210i&0.315-0.010i\\
-0.054+0.210i&0.340&-0.106+0.262i\\
0.315+0.010i&-0.106-0.262i&0.305
\end{array}  \right)
\label{eq:beta}
\end{equation}
\end{center}

    The eigenvalues of this matrix are $\lambda_{1}=0.877, \lambda_{2}=0.136, \lambda_{3}=-0.013$.
    A corresponding set of eigenvectors is $X = (0.587,-0.173+0.521i,0.594-0.071i); Y =(0.642,0.379-0.649i,0.048+0.143i);
    Z =(0.493,-0.287+0.224i,-0.769-0.178i)$.
    Although the density matrix (Eq.~\ref{eq:beta}) is Hermitian and the condition $Tr(\rho)=1$ is satisfied, it doesn't correspond to any
    physical state because of the negativity of one of the eigenvalues. We want to point out
    that a first main component $(\rho_{exp}^{1})_{ij}=X_{i}X_{j}^{\ast}$ of a considered density matrix, which has a
    weight $0.878$ is already close to the theoretical state vector
    $|\beta^{\prime\prime}\rangle$ and the corresponding fidelity
    is $F=Tr(\rho_{th}\rho_{exp}^{1})=0.9903$.  The other two components correspond to the
    "experimental noise" that is due mainly to misalignments of a
    setup and small volume of collected data.
    Even at this point, the obtained raw fidelity values show the high quality of a generated state.
    We have obtained similar eigenvalues for all other states and raw fidelity computed for the main density matrix
    component as described above have varied
    from 0.983 to 0.998. We also employed the maximum likelihood method of quantum state root
    estimation (MLE) \cite{bogdan:03a,bogdan:03b} to make a tomographically reconstructed
    matrix satisfy its physical properties, such as positivity.
    The results are presented in the following table (Table II). The
    level of statistical fluctuations in fidelity estimation was
    determined by the finite size of registered events $(\sim 500)$.
    All experimental fidelity values lie within the theoretical range
    of $5\% (F=0.9842)$ and $95\% (F=0.9991)$ quantiles
    \cite{quant,bogdan:03a}.

\begin{figure}[!ht]
\includegraphics[width=0.25\textwidth]{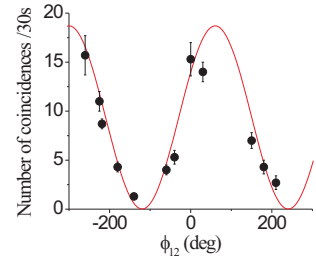}
\caption{Dependence of number of coincidences on a phase
$\phi_{12}$ for a given settings of polarization filters.}
\label{ort}
\end{figure}

\begin{center} \footnotesize
\begin{table}[!ht]
\begin{tabular}{|c|c|c||c|c|c|}\hline
State&$F_{MLE}$&State&$F_{MLE}$&State&$F_{MLE}$\\\hline
$|\alpha^\prime\rangle$&$0.9989$&$|\alpha^{\prime\prime}\rangle$&$0.9967$&$|\alpha^{\prime\prime\prime}\rangle$&$0.9883$\\
\hline
$|\beta^\prime\rangle$&$0.9967$&$|\beta^{\prime\prime}\rangle$&$0.9989$&$|\beta^{\prime\prime\prime}\rangle$&$0.9989$\\
\hline
$|\gamma^\prime\rangle$&$0.9883$&$|\gamma^{\prime\prime}\rangle$&$0.9883$&$|\gamma^{\prime\prime\prime}\rangle$&$0.9967$\\
\hline
\end{tabular}
\caption{Fidelities estimated with Maximum Likelihood Method}
\end{table}
\end{center}
The obtained fidelity values show the high quality of the prepared
states. Altogether with the high visibility of the interference
patterns that we obtained, we can conclude that our technique for
the preparation of qutrits is quite reliable.  The other test of
the quality of prepared states is the fulfillment of the
orthogonality criterion for the states that belong to the same
basis. For each set state we calculated the settings of waveplates
in our measurement setup that ensured the maximal projection of
each photon on the vertical polarization direction. In
Fig.~\ref{ort} we show the dependence of the coincidence rate for
the following setting of waveplates
$\chi_1=28.3^\circ,\theta_1=-33.5^\circ,\chi_2=-24^\circ,\theta_2=-2^\circ$.
These values correspond to the set state
$|\alpha^{\prime\prime\prime}\rangle$. As one can see, for the
fixed value $\phi_{13}=0$ the coincidence rate is almost equal to
zero, when phase $\phi_{12}=-120^\circ$. This corresponds to the
generation of the state $|\beta^{\prime\prime\prime}\rangle$,
which is orthogonal to $|\alpha^{\prime\prime\prime}\rangle$. The
visibility of this pattern is equal to 93.2\%. For the other
bases, the obtained values of visibilities varied from 92\% to
95\%. With these values of visibility, the lowest value of
coincidence rate corresponds to the accidental (Poissonian)
coincidence level and therefore the obtained data verifies the
orthogonality criterion.

\textbf{Conclusions}. We realized an interferometric method of
preparing the three-level quantum optical systems, that relied on
the polarization properties of single-mode two-photon light. The
specific sequence of states was generated and measured with high
fidelity values. The orthogonality of the states that form
mutually unbiased bases was experimentally verified. As an
advantage of this method we note that all control of the
amplitudes and phases of each basic state in superposition
(\ref{eq:state}) is done using linear optical elements, making it
easy to switch from one state to another and providing the full
control over the state (\ref{eq:state}). The main disadvantage is
that we cannot generate an entangled qutrits in this
configuration. Our setup also allows one to prepare an arbitrary
polarization qutrit state on demand.

\textbf{Acknowledgments}. Useful discussions with A.Ekert,
B.Englert, D.Kazlikowski, C.Kurtsiefer, L.C.Kwek, A.Lamas-Linares
and A.Penin are gratefully acknowledged. This work was supported
in part by Russian Foundation of Basic Research (projects
03-02-16444 and 02-02-16843) and the National University of
Singapore's Eastern Europe Research Scientist and Student
Programme.

\end{document}